 \documentclass[prl,aps,twocolumn,showpacs]{revtex4}
\usepackage[dvips]{graphicx}
\usepackage{dcolumn}%
\usepackage{amsmath}%
\usepackage{amsfonts}%
\usepackage{amssymb}
 \usepackage[usenames]{color}
\providecommand{\U}[1]{\protect\rule{.1in}{.1in}}
\newcommand{\be}{\begin{equation}}
\newcommand{\ee}{\end{equation}}
\newcommand{\bea}{\begin{eqnarray}}
\newcommand{\eea}{\end{eqnarray}}
\newcommand{\bt} {\begin{tabular}}
\newcommand{\et} {\end{tabular}}
\newcommand{\nn}{ \nonumber}

\newcommand{\ba} {\begin{array}}
\newcommand{\ea} {\end{array}}
\begin{document}

\title{Electron transport mechanisms in polymer-carbon sphere composites}

\author{  Cesar A. Nieves, Idalia Ramos,   Nicholas J. Pinto, and Natalya A. Zimbovskaya}

\affiliation
{Department of Physics and Electronics, University of Puerto 
Rico,  Humacao, Puerto Rico 00791, USA}

\begin{abstract}
A set of uniform carbon microspheres (CS) whose diameters have the order of $ 0.125\ \mu m$  to $10 \ \mu m $ was prepared from aqueous sucrose solution by means of hydrothermal carbonization of sugar molecules. A pressed pellet was composed by mixing CSs with polyethylene oxide (PEO). Electrical characterization of the pellet was carried out showing Ohmic current-voltage characteristics and temperature-dependent conductivity in the range $ 80\ K < T < 300\ K.$ The conductivity reached a maximum value of $ 0.245\ S/cm $ at $ 258\ K. $
 The dependence of conductivity on temperature was theoretically analyzed to determine predominating mechanisms of electron transport. It was shown that thermally-induced electron tunneling between adjacent spheres may take on an important part in the electron transport through the CS/PEO composites.
  \vspace{2mm}
	
	
	\end{abstract}


\date{\today}
\maketitle

\section{i. introduction}

Carbon compounds exist in various forms, some of which are electrical insulators, some are electrical conductors and some are semiconducting. These materials exhibit a range of chemical, physical and mechanical properties depending on the bonding between carbon atoms and the resulting crystal structure. Common forms of carbon-based materials that have been studied before include spheres, tubes, sheets and fibers. Over the past two decades, the field of nanoscience and nanotechnology has focused on carbon based materials like fullerenes \cite{1,2}, carbon nanotubes \cite{3} and graphene \cite{4} due to their promise to revolutionize the electronics industry. Recently however, there has been renewed interest in carbon microspheres for use in super strength carbon based objects and passive electronic components such as supercapacitors 
\cite{5,6}. Various techniques have been discovered to make these spheres which include chemical vapor deposition, templates, pyrolysis of carbon rich sources and hydrothermal carbonization (HTC) of large sugar molecules \cite{7,8,9,10,11,12,13,14}. This interest results from the easy one pot hydrothermal method of preparation that yields in large quantities uniform carbon spheres (CS) that exhibit high conductivity via post treatment and are chemically very stable under standard atmospheric conditions.
In this work, we present our results on the CSs conductivity $ (\sigma) $ as a function of temperature in the range $80\ K-300\ K. $ The electrical conductivity of a pressed pellet composed of CS and Poly(ethylene) Oxide (PEO) was  $ 0.22\ S/cm $ at room temperature and dropped to $ 0.18\ S/cm $ at $ 80\ K, $ passing through a maximum of $ 0.24\ S/cm $ at $ 258\ K. $ The PEO was used as a platicizer in the pellet preparation. 

We  analyze possible contributions of various transport mechanisms to most plausibly interpret
experimental results. Carrying on this analysis we pay a special attention to thermally-induced electron tunneling between adjacent CSs. This transport mechanism was first suggested by Sheng  \cite{15} to describe the tunneling conduction in disordered systems and later employed to analyze electron transport in various polymer/carbon composites. We show that this transport mechanism is mostly responsible for the maximum in the  temperature dependence of observed CS/PEO sample conductivity, whereas the thermally-activated transport predominates at lower temperatures. A unique feature of the presented model  is that it covers the entire temperature range from  $ 80\ K $ to $ 300\ K.$

\section{ii.  experiment}

A schematic of the preparation method of the CSs is shown in Fig. 1. A $ 0.8\ M $ aqueous sucrose solution in distilled water was transferred into a $ 50\ mL $ stainless-steel autoclave. The solution was heated at $200^o\ C $ for 4 hours and then cooled down to room temperature. The sucrose molecules underwent hydrolysis giving rise to monosaccharides  such as glucose \cite{16}. Subsequent dehydration of the monosacharides promoted nuclei-oligomeres formation within spherical micelles. Further polymerization led to the increment in the spheres diameters determining the spheres size. Additional carbonization has little effect on the CSs morphology, as  demonstrated in Ref.  \cite{7}. After cooling the solution
     the black precipitate was collected and washed with water and ethanol to remove residues from the sucrose decomposition and then dried at $70^o\ C $  for 12 hours. The as-prepared spheres were thermally annealed in a tube furnace under flowing nitrogen at  $ 800^o\ C.$ The sample was heated at a rate of  $10^o\ C/min $  until $ 800^o\ C. $ This temperature was held for 1 hour and then allowed to cool down. 
     
\begin{figure}[t] 
\begin{center}
\includegraphics[width=6cm,height=6.5cm,angle=-90]{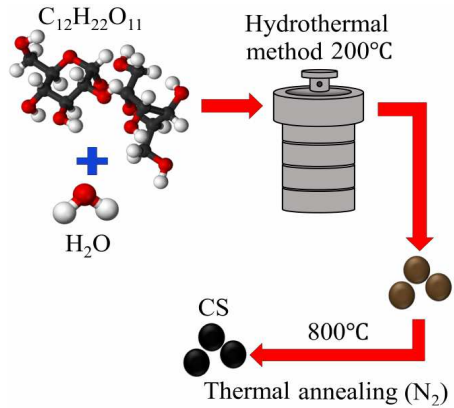} 
\caption{(Color online) Schematic of hydrothermal method and thermal annealing process used to produce carbon spheres with sucrose as the carbon source.
}
 \label{rateI}
\end{center}\end{figure} 
     
     A JEOL JMS-6360 Scanning Electron Microscopy (SEM) was used to characterize the morphologies and size distributions of the spheres. Energy Dispersive Spectroscopy (EDS) was used to analyze the chemical composition. The structural characterization was performed by X-Ray Diffraction (XRD) on a Bruker D2-Phaser XRD with a $  CuK_\alpha  = 0.154\ nm $ source. 
To fabricate the pellet, we mixed CS and PEO (MW 200,000, Sigma Aldrich) in a $ 1:1 $ mass ratio and ground the mixture in a mortar and pestle. The pellet consists of nearly homogeneous dispersion of CSs and PEO which was used as the adhesive agent. However, our studies of the pellet electrical conductivity give grounds to suggest that some CSs are probably arranged in chains extended through the sample. This arrangement does not destroy the general dispersion homogeneity provided that a relatively small amount of CSs are included into such chains. Without PEO, the pressed pellet was difficult to manipulate as it would easily crack making it impossible to measure its conductivity.  We aimed at preparation of a manageable sample with high concentration of CSs. We have found that a pellet characterized by $1 : 1 $ mass ratio is sufficiently hard whereas pressed pellets with higher CSs concentrations appeared too brittle to handle. Therefore, we proceeded with experiments using this ratio. The corresponding volume concentration of PEO in the sample was estimated to be about 40\%. Since the sample was electrically conducting, we believe that the content of carbon spheres was above the percolation threshold.  The CS/PEO sample was compressed using a Carver Press (Wabash, IN) under a load of 20,000 lbs and cut into a rectangular shape. Silver glue was used to fabricate two metallic electrodes attached to the sample to perform  electrical measurements. The current-voltage $ (I-V) $ characteristics of the pellet were measured at $300\ K $ using a Keithley $ 6517\ A $ Electrometer and a Keithley 6487 picoammeter/voltage source in a vacuum of $ \sim 10^{-2}\  Torr. $ A closed cycle helium refrigerator was used to lower the sample temperature which was controlled by a CryoCon 32 B temperature controller.  The pellet was mounted on a sapphire substrate ($Al_2O_3$ with $<0001> $ orientation and 0.5 mm thickness). Saphire is an electrical insulator and a good thermal conductor. The pellet was attached to the sapphire substrate using Apiezon N grease and the substrate was attached to the cold finger of the helium refrigerator with silver epoxy. The current in the pellet was measured at a fixed voltage of  $ 0.05\ V $ in the range of $ 300\ K$  to $ 80\ K $ from which the conductivity was later calculated.

The SEM images in Fig. 2(a) and Fig. 2(b) show low and high magnification images of typical CSs produced with HTC and thermal annealing. The CSs have spherical shapes with smooth surfaces and diameters ranging from $ 0.125\ \mu m $ to  $10\ \mu m.$ Further control of the sphere diameters can be obtained by varying the sucrose concentration in the precursor, as well as the temperatures and times for HTC and thermal annealing \cite{13}. The EDS spectra in Fig. 2(c) confirm that the spheres are composed of 99.5\% carbon. The XRD pattern in Fig. 2(d) has characteristic peaks at $22.5^o $  and  $ 44^o $, corresponding to reflections of 002 and 101 planes of graphitized carbon \cite{10}. 

\begin{figure}[t] 
\begin{center}
\includegraphics[width=8cm,height=9cm,angle=-90]{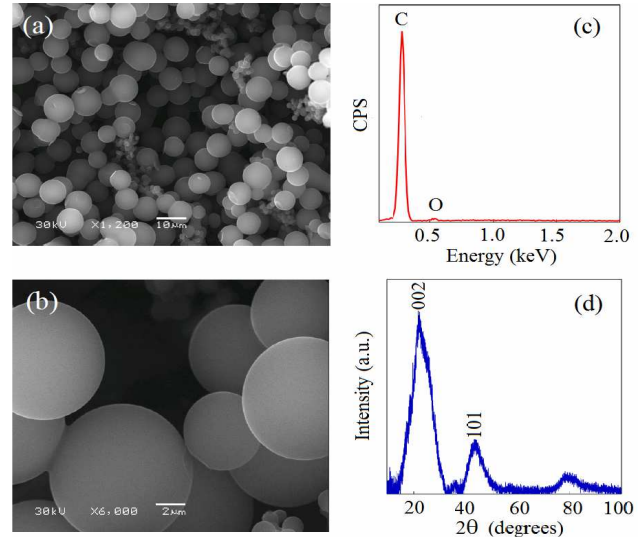} 
\caption{(Color online) SEM images: (a) low magnification $(10\ \mu m $ scale),  (b) high  magnification $(2 \ \mu m $ scale), (c)  EDS spectra showing corresponding CS composition and (d) XRD pattern of CS.  
}
 \label{rateI}
\end{center}\end{figure} 

The thermal annealing process eliminates  functional groups and increases the crystallinity of the CSs in agreement with reports in literature \cite{17}. These physical characteristics greatly influence the conductivity of the carbon spheres. Fig. 3(a) shows  plots of the current-voltage (I-V) curves of the CS/PEO pellet taken at $300\ K $ and $ 80\ K .$  The schematic diagram in the inset of the figure demonstrates Ag electrodes used for the electrical characterization and a SEM image of a section of the pellet. The spheres retain their shape even under compression. The I-V curves  indicate an Ohmic response and the absence of Schottky contacts between the Ag electrodes and the CS/PEO. The conductivity of the CS/PEO  was calculated from Fig. 3(a) to be $ 0.22\ S/cm $  at room temperature and $ 0.18\ S/cm $ at $ 80\ K .$. The temperature dependence of the conductivity shown in Fig.  3(b) shows a drop to $ 0.18\ S/cm $ when the temperature decreases from room temperature to $ 80\ K. $  

\begin{figure}[t] 
\begin{center}
\includegraphics[width=4.5cm,height=9cm,angle=-90]{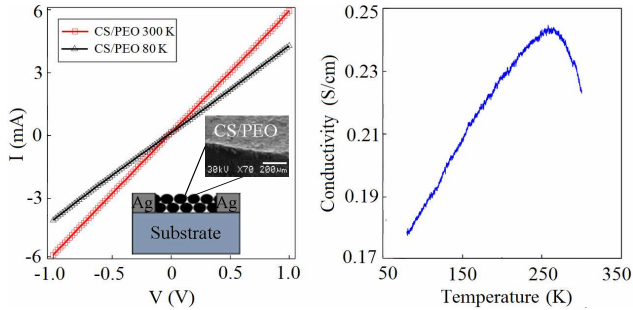} 
\caption{(Color online) Left panel: Current-voltage (I-V) characteristics curve of the CS/PEO composite showing Ohmic behavior at room temperature and $ 80\ K. $ Upper Inset: Schematic of the CS/PEO device and the corresponding SEM image of the composite. Right panel: Temperature-dependent conductivity of the CS/PEO composite.
}
 \label{rateI}
\end{center}\end{figure} 

Our results also indicate a change in the slope, with the conductivity increasing as temperature is lowered from $ 300\ K $ to $ 258\ K $ and then decreasing as temperature is further lowered down to $ 80\ K. $ Similar behavior of temperature-dependent conductivity  was previously reported for some polymer/carbon nanotube composites \cite{18,19,20,21}, polymer/carbon fibers composites \cite{22,23,24} and polymer/carbon black composites \cite{25}. In each case, the specifics of conductivity/resistivity temperature dependence is governed by the interplay between several different transport mechanisms. Relevant transport mechanisms include thermally-activated electron transport of the same kind as that predominating in intrinsic semiconductors, variable range electron hopping (VRH) and electron tunneling. In the following section we  analyze contributions of these mechanisms to electron transport in CS/PEO composites basing on the above described experimental results.

\section{iii. Discussion}

Analyzing electron transport mechanisms in CS/PEO,
it is reasonable to assume that some of the spheres are stuck together and arranged in chains extended through the whole sample thus providing pathways for traveling electrons  similar to those observed in self-assemblies of monodisperse starbust CSs \cite{5}.
This assumption agrees with experimental data shown in Fig. 4. 
    The plot  of $ \ln\sigma $ versus $ 100/T $ displayed in the left panel of the figure includes a nearly rectilinear piece corresponding to lower temperatures $(T \sim 80-110\ K). $ So, within this temperature range the   conductivity $ \sigma $ mostly   originates from a process of thermal activation typical for semiconducting materials and the effect of other transport mechanisms is negligible. The corresponding  conductivity $ \sigma_1, $ can be expressed by a standard Arrhenius relation:
\be
 \sigma_1 = \sigma_{10} \exp \big [-\Delta E/kT\big].                 \label{1}
\ee
Here, the prefactor $ \sigma_{10} $ does not depend on temperature, $ \Delta E $ is the activation energy and $ k $ is the Boltzmann constant. Basing on the data presented in Fig. 4, one concludes that for the considered sample $ \Delta E \approx 2.34\ meV, $ which nearly coincides with the estimation made for carbon fibers in  polyethylene-carbon fiber composites $(\Delta E \approx 2.4\ meV) ,$ and is reasonably close to the activation energy for networked-nanographite   $( \Delta E \approx 4.0\ meV) $ \cite{24}. 
At higher temperatures, the $\ln \sigma $ vs $ 100/T $ plot departs from the straight-line profile. This indicates that other mechanisms for electron transport start to noticeably contribute to the net conductivity although thermally activated transport still predominates.  The appearance of distinguishable signatures of  transport mechanisms different from thermal activation may be explained assuming
   that  in the considered sample the most of the spheres are separated from each other, and only the minority is combined in chains free from gaps between the spheres. This assumption agrees with SEM images shown in Fig. 2(a,b).

\begin{figure}[t] 
\begin{center}
\includegraphics[width=8.8cm,height=4.5cm]{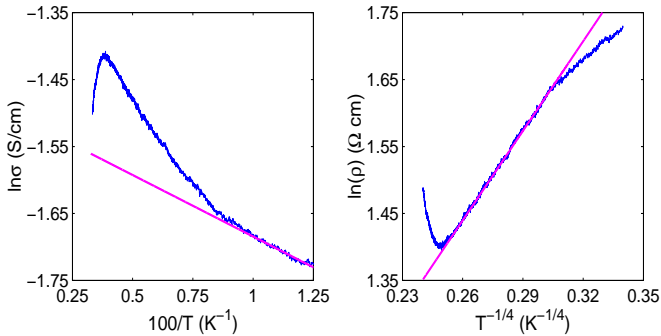} 
\caption{(Color online)  Logarithmic plot  of the CS/PEO composite conductivity versus $ T^{-1} $ (left panel) and logarithmic plot of the sample resistivity versus $ T^{-1/4} $ (right panel).
}
 \label{rateI}
\end{center}\end{figure} 
 
Electron transport through the gaps separating adjacent spheres may be controlled by two different transport mechanisms. These are the variable range hopping of electrons (VRH) between the sites on PEO polymeric chains filling the gaps between the spheres and the electron tunnelling through the gaps. The VRH transport theory was first suggested by Mott \cite{26}. Mott's original theory was later applied to analyze transport mechanisms in noncrystalline substances, doped semiconductors and polymers. When the electron transport is governed by the VRH, the resistivity is given by:
\be
\rho = \rho_0 \exp \big[(T_h/T)^{1/4} \big].  \label{2}
\ee
Here, $ \rho_0 $ is temperature-independent, and the characteristic temperature $ T_h $ has the form:
\be
T_h = 16 \big/ kN(E_F) l^3.   \label{3}
\ee
As follows from this expression, $ T_h $ is inversely proportional to the density of localized electron states on the Fermi level $N(E_F).$ Also, it strongly depends on the localization length of electrons $ l. $ The latter is directly proportional to the mean hopping distance $ R $ \cite{27}:
\be
R = \frac{3}{8} l \left(\frac{T_h}{T}\right)^{1/4}.  \label{4}
\ee

Usually, the VRH theory is employed at $ T \ll T_h $ when $ R $ significantly exceeds $ l. $ It is considered as not applicable when $ T $ is greater than $ T_h $ and $ R $ takes on values smaller than the localization length. To consider a possible contribution of VRH into electron transport characteristics observed in the present experiments we analyze the plot of $ \ln\rho $ versus $ T^{-1/4} $ shown in the right panel of Fig. 4. In this plot one may separate out two nearly rectilinear pieces with the crossover at $ T \approx 100\ K. $ As discussed before, the thermally-activated electron transport through chains of glued together carbon spheres is likely to determine the temperature dependence of conductivity/resistivity of the sample at $ T < 100\ K. $ So, the closeness of the corresponding part of the plot to a straight line seems to appear by accident. It probably occurs due to a lower sensitivity of $ T^{-1/4} $ plots compared to  $ T^{-1} $ plots. At higher temperatures $(T > 100\ K)$ the thermally-activated conductivity becomes nearly independent of temperature because of the low value of the activation energy, so it cannot determine the temperature dependence of conductivity/resistivity. From the slope of the corresponding straight line we find the characteristic VRH temperature $ T_h \approx 410\ K. $ Then we use Eqs. (\ref{3}) and (\ref{4}) to estimate the localization length and the mean hopping distance. Assuming that in PEO $ N(E_F) \approx 2.4\cdot 10^{21}\ states/eVcm^3 $ \cite{28} we obtain $ l \approx 5.6\ nm $ and $ R \approx 2.2-2.6\ nm $ within the temperature range $ 100\ K < T < 200\ K. $ Since $ R $ is significantly smaller than $ l, $ we may presume that VRH mechanism does not strongly contribute to the sample conductivity. At the same time, a comparatively large value of localization length gives grounds to expect that electron transport between the carbon spheres is determined by electron tunneling through potential barriers.

Analyzing electron tunneling conductivity we remark that the tunnel current may be strongly affected by voltage fluctuations across the gaps as suggested by Sheng \cite{15}. In the considered case, these fluctuations originate from thermally excited motions of ions on the surfaces of adjacent spheres. They  distort profiles of energy barriers separating the spheres and cause temperature dependences of tunnel conductivity. The effect of thermally induced voltage fluctuations on the tunneling conduction was taken into account in studies of electron tunnel transport in various polymer-carbon composites (see e.g. Refs. \cite{18,19,20,23,24}).

Within the Sheng's theory, the effect of thermal fluctuations is described by introducing an extra electric potential $ V_T $   associated with fluctuations. A tunneling electron sees a potential barrier whose original rectangular profile is modified by the image-force correction and further distorted by thermal fluctuations.  The barrier width is determined by the minimum distance between adjacent spheres $ d $ for the tunneling probability exponentially decreases as the tunneling distance rises. The simplest barrier profile is given by the expression:
\be
V(u,\epsilon) = V_0 \big[(1 - u) u - \epsilon u\big].  \label{5}
\ee
Here, $ x $-axis is supposed to run across the barrier $(0\leq x \leq d),\ u = x/d,\ \epsilon = V_T/V_0 $ and $ V_0 $ characterizes the  height of the rectangular  barrier free from the image-force correction. The maximum height of the barrier $ \big [V_m  = \frac{V_0}{4}(1 - \epsilon)^2 \big] $  is reached at $ u = \frac{1}{2}(1 - \epsilon). $ The expression (\ref{5}) implies that at low temperatures when one may omit the effect of thermal fluctuations $(\epsilon \to 0 ) $ the barrier's profile acquires   a parabolic lineshape. The parabolic profile distortion by thermal fluctuations is illustrated in Fig. 5. Simmons \cite{29} and Sheng \cite{15} showed that the potential barrier's profiles may be modeled by a wide variety of lineshapes. Nevertheless, the chosen parabolic approximation was (and still is) used to explain experimental results concerning  transport in carbon-polymer composites (see e.g. Refs. \cite{18,19,20,23}).

\begin{figure}[t] 
\begin{center}
\includegraphics[width=8.8cm,height=4.5cm]{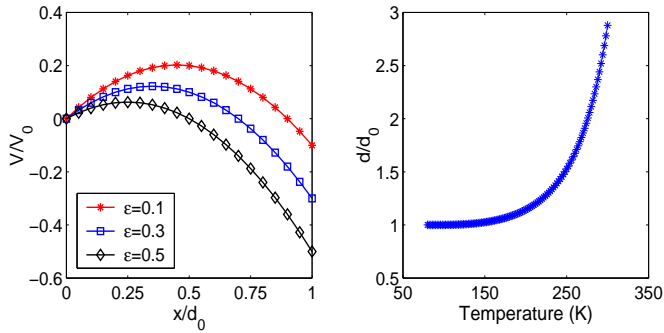} 
\caption{(Color online)  Calculated characteristics of the potential barrier separating two adjacent carbon spheres. Left panel: The effect of thermally induced voltage fluctuations on the barrier profile. Right panel: The average barrier width as a function of temperature.  $(d_0 $ is corresponding to  the barrier width at $ T = 80\ K). $
}
 \label{rateI}
\end{center}\end{figure}

The considered  composite necessarily contains a large number of tunnel junctions with diverse  characteristics, and the effects of voltage fluctuations vary from one junction to another. The set of tunnel junctions may be regarded as random resistor network where resistors are identified with sole junctions. The network conductivity/resistivity can be calculated by the self-consistent effective-medium theory \cite{30}. Within this approach, the average effect of random conductances associated with the junctions may be treated by replacing all of them by a single effective conductance, so the complete network is reduced to a single junction. 
 It was suggested \cite{15} to treat the most preferable region for electron tunnelling through the junction as a parallel plate capacitor. In the considered case, when the tunnelling occurs between adjacent CSs one may identify the separation between the capacitor plates with the minimum distance between the spheres $ d$ and approximate the plate area as $\pi d^2/4.$  Then the capacitance of this capacitor may be estimated as $C \approx \epsilon_0 \kappa\pi d,\ \kappa $ being the PEO dielectric constant.

The Sheng's model is analytically solvable provided that the barrier profile is approximated by Eq. (\ref{5}). Carrying out calculations described in the Appendix one obtains the following expression for the thermally-induced electrical conductivity:
\be
\sigma_2 = \sigma_{02} \left(1 + \frac{T}{T_0}\right)^{-3/2} \exp\left[- \frac{T_1}{T + T_0} \right].   \label{6}
\ee
Here, the prefactor $ \sigma_{02}: $ 
\be
\sigma_{02} = \frac{me^2}{(2\pi\hbar)^3} \frac{eV_0}{\chi},    \label{7}
\ee
and the tunneling constant $ \chi: $
\be
\chi = \sqrt{\frac{2meV_0}{\hbar^2} }            \label{8}
\ee
are temperature-independent, and two characteristic temperatures $ T_0 $ and $ T_1 $ are determined by the relation:
\be
kT_1 = \frac{CV_0^2}{2} = \frac{\pi d\chi}{4}kT_0.   \label{9}
\ee
where $ C $ is the capacitance of a parallel plate capacitor representing the region where the tunneling between the spheres mostly occurs.
As $ \chi d> 1, $ the temperature $ T_0 $ is lower than $ T_1. $ At low temperatures $(T\ll T_0), $ the conductivity $ \sigma_2 $ is reduced to a simple tunneling conductivity described by a well known expression: $ \sigma_2 = \sigma_{02} \exp [-\pi d\chi/4). $ 
  As the temperature rises, the conductivity $ \sigma_2 $ increases and it reaches its maximum value at $ T_m = \frac{2}{3} T_1 - T_0 $ provided that $ \chi d $ is sufficiently large  $ (\chi d > 6/\pi). $ When the temperature grows above $ T_m, \ \sigma_2 $ decreases. This conductivity behavior 
qualitatively  agrees with that observed in the experiment.

We remark  that both thermally activated transport through chains of carbon spheres and VRH transport mechanism are associated with conductivities which smoothly increase as the temperature grows and do not show any maxima/minima at certain values of $ T. $ This gives grounds to expect the thermally induced electron tunneling between the carbon spheres to significantly contribute  to electron transport in the considered system. Also, we remark that the present result for the thermally induced tunnel conductivity is obtained by straightforward computations of all relevant integrals. 
It is more accurate than the corresponding expressions used in earlier works (see e.g. Ref. \cite{24}) where the power factor $ (1 + T/T_0)^{-3/2} $ was lost. This power factor is responsible for the appearance of maximum in  $ \sigma_2 $ versus $ T $ curves, so the difference between our Eq. (\ref{6}) and earlier results  is important and may not be disregarded.

Basing on the above described analysis we conclude that electron transport in the considered CS/PEO composite is mostly governed by two transport mechanisms, namely, thermally-activated transport through chains compiled of adjoining spheres and thermally-induced tunneling between the spheres separated by gaps. Correspondingly, the net conductivity may be presented in the form:
\be
\sigma = \sigma_1 + \sigma_2    \label{10}
\ee
where $ \sigma_ 1 $ and $ \sigma_2 $ are given by Eqs. (\ref{1}) and (\ref{6}), respectively. The net conductivity still shows a maximum at a certain temperature. However, the maximum  position is shifted with respect to $ T_m $ due to the effect of the term $ \sigma_1. $

Using the expression (\ref{10}), one may reach a reasonably good fit between theoretical and experimental results. The fit may be further improved if one more factor is taken into account.  As discussed in Ref. \cite{24}, the temperature growth leads to thermal expansion of air voids in PEO and to intensification of mobility of polymeric chains originating from their macro-Brownian motion. 
As a result, the minimum distances between the spheres increase as the temperature rises. 
The temperature dependence of $ d $ was determined by a computer simulation. The simulation results showed that the best fit of theoretical results to the experimental data could be achieved using a nearly constant value of $ d $ at temperatures below 150 K replaced by its rather fast increasing  at higher temperatures.
   This is illustrated in the right panel of Fig. 5.  Also, it was shown that the best fitting values for the temperatures $ T_0 $ and $ T_1 $ equal $200\ K $ and $1850\ K,$ respectively.
        Temperature dependence of conductivity given by Eq. (\ref{10}) superimposed over the experimental data is shown in Fig. 6. A good agreement between the theory and experiment is demonstrated.  In plotting of theoretical curve it was  assumed that $ d $ varies with the temperature, as shown in Fig. 5.   
 The behavior of $ d $ presented in the figure is responsible for making the maximum in the conductivity more sharp and distinct. The curve plotted in Fig. 5 may be approximately represented by an analytical expression: 
\be
d \approx 0.5d_0 \Big\{1 + \exp\Big[-0.075 \big(1 - {T}/{T_1} \big)^3 \Big]\Big\}   \label{11}
\ee  where $T_1 = 1850\ K$.  
  To further examine the applicability of Eq. (\ref{10})  we estimate some relevant parameters characterizing the thermally-induced tunneling electron transport.

\begin{figure}[t] 
\begin{center}
\includegraphics[width=8.8cm,height=4.5cm]{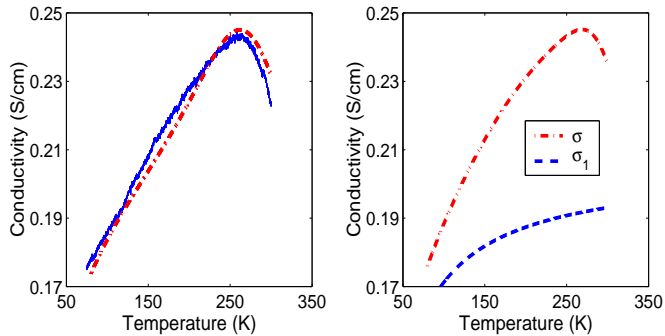} 
\caption{(Color online)  left panel: Experimental (solid line) and theoretical (dash-dotted line) results representing the temperature dependence of the conductivity of the CS/PEO composite. The theoretical result is computed using Eqs. (\ref{1}) and (\ref{6}) and assuming that $ T_1 =  1850 K,\ T_0 = 200\ K,\ \sigma_{10} = 0.205\ S/cm,\  \sigma_{20} = 11\ S/cm $ and  $ \Delta E = 2.34\ meV.$   Right panel:  The contribution of thermally activated transport through the chains of adjoining carbon spheres $(\sigma_1) $ to the net conductivity of the sample $ (\sigma)$. The curves are plotted using the same values of relevant parameters as before.
}
 \label{rateI}
\end{center}\end{figure}

As was mentioned before, the diameters of carbon spheres in the sample  $(D) $ vary between $ 0.125\ \mu m$ and $ 10\ \mu m,  $ the mean and median diameters being equal  $ 2.93\ \mu m $ and $ 2.0\ \mu m, $ respectively. Accordingly, in further estimations we choose $ D \sim 2\ \mu m. $ 
 The spheres occupy $ \sim 40\% $ of the sample volume. So, one may estimate minimum distances between adjacent spheres $ d \sim 10 - 100\ nm. $ The PEO dielectric constant $ \kappa $ equals  $ 2.24. $ Using these data to estimate the capacitance $ C $ and substituting the result into Eq. (\ref{9}) one approximates $ V_0 $ as $ 0.3 - 0.1\ V. $  The temperature dependence of conductivity was explored when the bias voltage $ V_b = 0.05\ V $ was applied across the sample. Therefore one may reasonably assume that bias voltage applied across a sole tunnel junction is much smaller than $ V_0. $ Also, the parameter $ \epsilon $ may be estimated as $ \frac{1}{V_0} \sqrt {kT/C} $ which follows from  the equipartition theorem. In the considered case, $ \epsilon $ varies between 0.18 and 0.3 when the temperature grows from $ 80\ K $ to $ 300 K ,$ so it remains significantly smaller  than $ 1 $ within the whole temperature range explored in the experiment. Finally, computing the ratio $ T_1/T $ we find that $ \chi d \approx 12. $ 
   Comparison of two terms in the Eq. (\ref{10}) shows that $ \sigma_1 $ exceeds $ \sigma_2 $ especially at low temperatures (see Fig. 6).  The contribution from the thermally-induced tunneling becomes more significant as the temperature rises. Moreover, as explained above, this contribution is responsible for the appearance of a maximum in the temperature dependence of conductivity.

\section{iv. conclusion}

Presently, carbon microspheres attract a significant attention due to their potentialities in building up elements for microelectronic devices such as supercapacitors. We prepared a large number of uniform CSs with diverse  diameters from an aqueous sucrose solution using hydrothermal carbonization of large sugar molecules.  The spheres were mixed with PEO resulting in fabrication of a CS/PEO composite. Electrical characterization of the sample did show an Ohmic response and the absence of Schottky barriers between the sample and the attached silver electrodes. The sample conductivity appeared to be rather high despite the effect of a non-conducting PEO filling gaps between CSs.

The conductivity is temperature-dependent. As the temperature is lowered down from $300K $ to $ 80K, $ the conductivity first increases and reaches the maximum value at $ T \approx 258 K $ and then it decreases following the temperature decrease. We have analyzed transport mechanisms which may control electron transport in the CS/PEO sample. Our analysis gives grounds to suggest that there exist some chains of glued together CSs extended through the whole sample. These chains form a network of channels for thermally-activated electron transport typical for intrinsic semiconductors. We remark that our estimate for the activation energy is reasonably close  to the value obtained for networked-nanographite which confirms the above suggestion. As follows from experimental results, thermally-activated transport  predominates  within the whole temperature range, especially
 at lower temperatures. 

Another important transport mechanism is electron tunneling between adjacent CSs affected by thermally-induced voltage fluctuations distorting profiles of energy barriers separating the spheres. Using a parabolic approximation for the energy barrier separating adjacent CSs we have derived a novel expression for thermally-induced conductivity whose temperature dependence shows a maximum at a certain temperature.
 On the contrary, thermally-activated electron transport as well as VRH leads to conductivity which smoothly decreases  following the decrease in temperature. Our analysis shows that thermally- induced electron tunneling probably brings a significant contribution to the sample conductivity at temperatures close to/or higher than one corresponding to the conductivity maximum. Also, we have studied possible contribution from VRH transport mechanism. Our analysis showed that this mechanism is unlikely to significantly contribute to the temperature-dependent conductivity of the considered CS/PEO sample. This conclusion by no means disagrees with the importance of VRH in general.

Finally, we conclude that the behavior of conductivity in the considered CS/PEO composite is mostly determined by the combined effect of thermally-activated electron transport and of thermally-induced electron tunneling between adjacent spheres. We believe that our results may contribute to further understanding of transport characteristics of polymer-carbon composites.
\vspace{4mm}

 {\bf Acknowledgments:}
 This work was supported  by  NSF-DMR-PREM 1523463.
We  thank  A. Melendez for providing us with SEM images and G. M. Zimbovsky for help with the manuscript.

\section{Appendix}

When fluctuating voltage $ V_T $  is much stronger than the bias voltage  $V_b $ applied across the junction, a partial conductivity may be defined as \cite{15}:
\be
\Sigma_T (V_T) = \frac{1}{2}\lim_{E_{b \to 0}} \frac{j(V_T + V_b) - j(V_T - V_b)}{E_b}.  \tag{A1}\label{11}
\ee
Here,  $E_b = V_B/d $ is the electric field corresponding to the bias voltage  across the junction and  $ j $ is the tunneling current density.

Fluctuation-induced thermal conductivity $ \sigma_2 $ should be obtained by averaging $ \Sigma_T $ with the fluctuation probability function $P(V_T): $
\be
\sigma_2 = \int_0^\infty P(V_T) \Sigma_T(V_T) d(V_T)  \tag{A2} \label{12}
\ee
where $P(V_T) $ has the form:
\be
P(V_T) = \sqrt{\frac{2C}{kT}} \exp \left[-\frac{CV_T^2}{kT} \right].   \tag{A3} \label{13}
\ee
In this expression, the capacitance $ C $ is associated with a parallel plate capacitor representing the most preferable region for tunneling between two carbon spheres. 

When the thermally  induced voltage exceeds the bias voltage across the tunnel junction, the tunneling current density is given by \cite{15}:
\begin{align}
j(V_T) = & \frac{mekT}{8\pi^2\hbar^3} \int_{-\infty}^\infty D(E,\epsilon)
\nn\\   & \times
\ln   \left\{\frac{1 + \exp [-E/kT]}{1 + \exp[-(E + eV_T)/kT]} \right\}dE  .      \tag{A4}     \label{14}
\end{align}
Here, $ m $ is the effective mass of a tunneling electron. It was shown \cite{31} that the barrier transmission factor $ D(E,\epsilon) $ for barriers with a slowly varying potential may be approximated as:
\be
D(E,\epsilon) = \left\{ \ba{ll}
\exp\big[-F(E,\epsilon)\big],  &\ E \leq V_m   \\
1, &\ E > V_m. 
\ea \right.   \tag{A5} \label{15}
\ee
The form of the function $ F(E,\epsilon) $ depends on the barrier profile. Using the adopted model of a parabolic barrier we get:
\be
F(E,\epsilon) = \frac{\pi d\chi}{4}\left[(1 - \epsilon)^2 - \frac{4E}{eV_0} \right].  \tag{A6} \label{16}
\ee
In this expression, $ \chi $ is the tunneling constant:
\be
\chi = \sqrt{\frac{2meV_0}{\hbar^2}}  \tag{A7}  \label{17}
\ee
which presumably takes on greater values than $ d^{-1}. $ Substituting expressions (\ref{15}),
(\ref{16}) into Eq. (\ref{14}) one may calculate the current density $j(V_T).$ As long as $ \epsilon $ takes on sufficiently small values $(\epsilon < 1),$  electron tunneling predominates over thermally-activated transport above the barrier, and the effect of temperature induced smearing of the Fermi level may be neglected. This brings explicit expressions   for the thermally-induced current density and the partial conductivity, namely:
\begin{align}
j(V_T) = &\frac{me}{8\pi^2\hbar^3} \left(\frac{eV_0}{\pi d\chi}\right)^2 \exp \left[-\frac{\pi d\chi}{4}(1-\epsilon)^2 \right]
\nn\\ & \times
\big\{ 1- \exp [-\pi d \chi \epsilon] \big\} \tag{A8}  \label{18}
\end{align}
\begin{align}
\Sigma_T(V_T) = & \frac{me^2}{2(2\pi\hbar)^3} \frac{eV_0}{\chi} \exp\left[-\frac{\pi d\chi}{4} (1 - \epsilon)^2 \right]
\nn\\ & \times
\big\{ 1+ \exp [-\pi d \chi \epsilon]  - \epsilon \big(1 - \exp[-\pi d\chi\epsilon]\big )\big\}  .\tag{A9}   \label{19}
\end{align}
Using Eqs. (\ref{12}) and (\ref{19}) we arrive at the result for the  thermally induced tunneling conductivity given by Eq. (\ref{6}).


\begin{thebibliography}{99}


 \bibitem{1}	S. Iijima, Direct observation of the tetrahedral bonding in graphitized carbon black by high resolution electron microscopy, J. Crystal Growth {\bf 50}, 675 (1980). 

 \bibitem{2}  H. W. Kroto, J. R. Heath, S. C. O'Brien, R. F. Curl, and R. E. Smalley, C60: Buckminsterfullerene, Nature {\bf 318}, 162 (1985).   

 \bibitem{3}	S. Iijima, Helical microtubules of graphitic carbon, Nature {\bf 354}, 56 (1991). 
             
 \bibitem{4} K. S. Novoselov, A. K. Geim, S. V. Morozov, D. Jiang, Y. Zhang, S. V. Dubonos, I. V. Grigorieva, A. A. Firsov,
Electric Field Effect in Atomically Thin Carbon Films, Science {\bf 306}, 666 (2004).  

 \bibitem{5}	S.-K. Kim, E. Jung, M. D. Goodman, K. S. Schweizer, N. Tatsuda, K. Yano, and P. V. Braun, Self-Assembly of Monodisperse Starburst Carbon Spheres into Hierarchically Organized Nanostructured Supercapacitor Electrodes, ACS Appl. Mat. \& Inter. {\bf 7}, 9128 (2015). 

 \bibitem{6}	Y. Jiang, M. Huang, X. Ju, and X. Meng, Fabrication of Hierarchical Porous Carbon Spheres for Electrochemical Capacitor Application, Chem. Lett. {\bf 45}, 48 (2016).  

 \bibitem{7}	Q. Wang, H. Li, L. Chen, and X. Huang, Monodispersed hard carbon spherules with uniform nanopores, Carbon {\bf 39}, 2211 (2001).  

 \bibitem{8}	Q. Wang, H. Li, L. Chen, and X. J. Huang, Novel spherical microporous carbon as anode material for Li-ion batteries, Solid State Ion. {\bf  43},  152 (2002).  

 \bibitem{9}	M.-M. Titirici, A. Thomas, and M. Antonietti, Replication and Coating of Silica Templates by Hydrothermal Carbonization, Adv. Func. Mat. {\bf 17}, 1010 (2007).  

 \bibitem{10}	M.-M.Titirici, A. Thomas, and M. Antonietti, Aminated hydrophilic ordered mesoporous carbons, J. Mat. Chem. {\bf 17}, 3412 (2007).  

 \bibitem{11}	Y. Mi, W. Hu, D. Youmeng, and Y. Liu, Synthesis of carbon micro-spheres by a glucose hydrothermal method, Mat. Lett. {\bf  62}, 1194 (2008).  

 \bibitem{12}	S. Tang, Y. Tang, S. Vongehr, X. Zhao, and X. Meng, Nanoporous carbon spheres and their application in dispersing silver nanoparticles, Appl. Surf. Sci. {\bf  255}, 6011 (2009).  

 \bibitem{13}	M. Li, W. Li,  and S. Liu, Hydrothermal synthesis, characterization, and KOH activation of carbon spheres from glucose, Carbohydrate Res. {\bf  346}, 999  (2011). 

 \bibitem{14}	J. Cao, Y. Wang, P. Xiao, Y. Chen, Y. Zhou, J.-H. Ouyang, and D. Jia, Hollow graphene spheres self-assembled from graphene oxide sheets by a one-step hydrothermal process, Carbon {\bf 56}, 383 (2013).  

\bibitem{15} P. Sheng, Fluctuation-induced tunneling conduction in disordered materials, Phys. Rev. B {\bf 21}, 2180 (1980).  

\bibitem{16}	M. Sevilla, A. B. Fuertes, Chemical and Structural Properties of Carbonaceous Products Obtained by Hydrothermal Carbonization of Saccharides, Chem.- A Eur. J.  {\bf 15}, 4195 (2009).  


\bibitem{17}	R.-R. Yao, D.-L. Zhao, L.-Z Bai, N. N. Yao, and L. Xu,  Facile synthesis and electrochemical performance of hollow graphene spheres as anode material for lithium batteries, Nanoscale Res. Lett. {\bf 9}, 368 (2014).

\bibitem{18} A. Nogales,  G. Broza,  Z. Roslaniec, K. Schulte, I. Sics, B. S. Hsiao, A. Sanz, M. C. Garcia-Gutierrez, D. R. Rueda, C. Domingo, and T. A. Ezquerra, Low Percolation Threshold in Nanocomposites Based on Oxidized Single Wall Carbon Nanotubes and Poly(butylene terephthalate)
Macromolecules   {\bf  37}, 7669 (2004).

 \bibitem{19} S. Barrau, P. Demont  A. Peigney,  C. Laurent, and C. Lacabanne, DC and AC Conductivity of Carbon Nanotubes-Polyepoxy Composites, Macromolecules   {\bf 36}, 5187 (2003). 

 \bibitem{20} A. Bello, E. Laredo, J. R. Marval, M. Grimau, M. L. Arnal,  A. J. Muller,  B. Ruelle, and P. Dubois, Universality and Percolation in Biodegradable Poly$(\epsilon-$caprolactone)/Multiwalled Carbon Nanotube Nanocomposites from Broad Band Alternating and Direct Current Conductivity at Various Temperatures, Macromolecules   {\bf  44}, 2819 (2011).
 
\bibitem{21} R. Zhang, M. Baxendale, and T. Peijs, Universal resistivity-strain dependence of carbon nanotube/polymer composites, Phys. Rev. B {\bf 76}, 195433 (2007).

 \bibitem{22} Y. Chekanov, R. Ohnogi, S. Asai, and M. Sumita,  Positive temperature coefficient effect of epoxy resin filled with short carbon fibers, Polymer J. {\bf 30}, 381 (1998).  

 \bibitem{23} D. Zhu, Y. Bin, and M. Matsuo, Electrical conducting behaviors in polymeric composites with carbonaceous fillers, J. Polymer Sci. B: Polymer Phys. {\bf 45}, 1037 (2007).
 
 \bibitem{24} R. Zhang, Y. Bin, R. Chen, and M. Matsuo, Evaluation by tunneling effect for the temperature-dependent electric conductivity of polymer-carbon fiber composites with visco-elastic properties, Polymer J. {\bf  45}, 1120 (2013).   

\bibitem{25} G. Wu, B. Li, and J. Jiang,  Carbon black 
self-networking induced co-continuity of immiscible polymer blends, Polymer, {\bf 51}, 2077 (2010).

 \bibitem{26} N. F. Mott and E. A. Davis, {\it Electronic processes in non-crystalline materials}, (Clarendon, Oxford, 1979).  
 
 \bibitem{27} R. Rosenbaum, Crossover from Mott to Efros-Shklovskii variable-range-hopping conductivity in In$_x$O$_y$ films,  Phys. Rev. B {\bf 44}, 3599 (1991). 
 
 \bibitem{28} M. Reghu, Y. Cao, D. Moses, and A. J. Heeger,  Counterion-induced processibility of polyaniline: Transport at the metal-insulator boundary, Phys. Rev. B {\bf 47}, 1758 (1993).

\bibitem{29}  J. G. Simmons, Generalized Formula for the Electric Tunnel Effect between Similar Electrodes Separated by a Thin Insulating Film,  J. Appl. Phys. {\bf 34}, 1793 (1963).  

\bibitem{30} S. Kirkpatrick, Percolation and Conduction, Rev. Mod. Phys. {\bf 45}, 574 (1973). 

 \bibitem{31} E. L. Murphy and R. H. Good, Jr., Thermionic Emission, Field Emission, and the Transition Region, Phys. Rev. {\bf 102}, 1484 (1956).   


 
 \end{thebibliography}
\end{document}